\documentclass[reprint,aps,pra,longbibliography]{revtex4-2}

\usepackage[T1]{fontenc}
\usepackage[utf8]{inputenc}
\usepackage{graphicx}
\usepackage{amsmath,amssymb}
\usepackage{siunitx}
\usepackage{hyperref}
\usepackage{orcidlink}
\usepackage{xcolor}
\colorlet{green}{green!60!black}
\usepackage{tikz}
\usetikzlibrary{calc,arrows.meta,decorations.pathmorphing,positioning}

\hypersetup{
  colorlinks=true,
  linkcolor=blue,
  citecolor=blue,
  urlcolor=blue
}

\usepackage{amsmath}
\usepackage{amssymb}
\usepackage{graphicx}
\usepackage{xcolor}
\usepackage{braket}      % For \ket, \braket
\usepackage{microtype}
\usepackage{bm}

\usepackage{pgfplots}
\pgfplotsset{compat=1.18}
\usepackage{tikz}

\DeclareMathOperator{\sign}{sign}

\newcommand{\rev}[1]{\textcolor{black}{#1}} % Referee 2 (blue)

\definecolor{darkgreen}{rgb}{0.0,0.35,0.0}
\newcommand{\revG}[1]{\textcolor{black}{#1}} % Referee 1 (dark green)

\newcommand{\revA}[1]{\textcolor{black}{#1}} % Author's further revision

\begin{document}

%------------------------------------------------------------------
% FRONT MATTER
%------------------------------------------------------------------
\title{Certified Private Relational Time from Entanglement}

\author{Karl Svozil}
\email{karl.svozil@tuwien.ac.at}
% ORCID: 0000-0001-6554-2802
\affiliation{%
    Institute for Theoretical Physics,
    TU Wien,
    Wiedner Hauptstra{\ss}e 8-10/136,
    1040 Vienna, Austria
}

\begin{abstract}
We introduce an ``entangled clock'' in which time is defined operationally by discrete measurement registrations on a singlet state. Locally, each party's tick rate is fixed by the unbiased marginals. The nontrivial resource is the relational (coincidence-tick) stream: because the singlet's information budget is entirely exhausted by joint properties, the only definite temporal structure resides in the correlations between the two parties. Operationally, after exchanging time tags and outcomes, Alice and Bob identify synchronized events (that is, the $++$ channel) and thereby obtain a joint tick record. Comparing the $++$ coincidence rate $R(\theta)=P_{++}(\vec a,\vec b)$ to Peres' isotropic bomb-fragment local-hidden-variable model (yielding $R_{\mathrm{cl}}(\theta)=\theta/(2\pi)$), we find that for obtuse analyzer separations the quantum prediction exceeds this natural classical benchmark, with a maximal relative excess of about $13.6\%$ near $\theta\approx 140.5^\circ$. We emphasize that this ``faster ticking'' refers to the \emph{rate of identified coincidence ticks} under a specific operational convention, not to an improved local clock rate, precision, or stability. Finally, by using multiple settings and a Bell test, we outline ``Certified Private Time'': a device-independent certification of unpredictability/privacy of the relational time-stamp record against adversaries lacking foreknowledge of the settings, analogous to certified randomness generation.
\end{abstract}


\maketitle

\section{Introduction}

The question of the fundamental nature of time and its measurement has recently been revisited through two distinct but complementary lenses: operational metrology in relativity and information-theoretic constraints in quantum foundations.

On the metrological side, the ``trialogue'' regarding the number of fundamental constants~\cite{DuffOkunVeneziano} established a consensus that dimensionless quantities are the ultimate observables. More recently, Matsas, Pleitez, Saa, and Vanzella~\cite{matsas-2024} argued that a relativistic spacetime requires only a single fundamental dimensional constant. By employing ``bona fide clocks,'' one can define the spacetime metric entirely through proper time measurements, rendering spatial rulers redundant.
This ``single-unit universe''---a cosmos whose metric geometry is fixed by
a single dimensional constant, namely proper time---relies on the mathematical rigidity of the Alexandrov-Zeeman theorems~\cite{alex3,zeeman}. These theorems state that the causal structure (light cones) determines the geometry of Minkowski spacetime up to a global conformal factor. The physical clock breaks this dilation symmetry, fixing the scale of the universe~\cite{svozil-2025-lt}.

If a single clock fixes the scale of the universe, a natural question arises: how do two such clocks, spatially separated but sharing a quantum history, measure the flow of time relative to one another? In classical relativity, synchronization is a matter of signaling and convention (Einstein synchronization~\cite{ein-05}). In quantum mechanics, by contrast, synchronization can be mediated by entanglement, which introduces nonclassical correlations between distant systems. These correlations are constrained by no-go theorems such as Bell's, and thus cannot in general be accounted for by local realistic statistics.

This brings us to the second lens: the study of quantum correlations. Zeilinger~\cite{zeil-99} proposed a foundational principle: that the irreducible randomness of quantum events arises because an elementary system carries only one bit of information. In an entangled state, this information is exhausted by the joint properties, leaving the individual constituents completely undefined. Simultaneously, Peres~\cite{peres222} demonstrated that classical correlations are strictly bounded because one can validly imagine the results of ``unperformed experiments.'' In contrast, quantum correlations defy these bounds precisely because such counterfactuals are illegitimate.

\revA{The time defined by the entangled clock is therefore fundamentally \emph{relational}: since the singlet state encodes no information about individual subsystem properties along any axis, the only definite temporal structure is contained in the \emph{joint} correlations between Alice's and Bob's measurement records. We call this ``relational time'' in the general sense that the relevant properties are bipartite/joint rather than marginal---without commitment to any specific interpretive program such as relational quantum mechanics~\cite{Rovelli1996}. Operationally, this relational structure manifests as a coincidence-tick stream in the $++$ channel after classical record exchange (Sec.~\ref{sec:coincidence-ticks}).}

\rev{A brief remark on context: there is a substantial literature on ``quantum clocks'' and time in quantum theory, ranging from early operational analyses of physical clocks and limits to time measurement to modern work on quantum time observables and quantum-enhanced metrology (see, e.g., Refs.~\cite{SaleckerWigner1958,Busch2008Time,Giovannetti2004,Moreva2014}). The present contribution is different in emphasis: rather than optimizing precision of time/phase estimation, we focus on an \emph{operational, Bell-theoretic} notion of a distributed clock in which the key figure of merit is an \emph{informational synchronization rate} (coincidence-tick density) and, in the multi-setting regime, device-independent certification of the unpredictability/privacy of the time-stamp record.}

\revG{A further clarification concerns the role of \emph{measurement} in the present construction. The protocol is formulated at the standard operational level of quantum theory: each run produces (i) a setting choice, (ii) a binary detector outcome, and (iii) a classical time tag. We deliberately do \emph{not} attempt to resolve the quantum measurement problem here (e.g., collapse vs.\ many-worlds vs.\ objective collapse). Instead, we take the experimentally standard viewpoint that a ``click'' constitutes a stable macroscopic record, well modeled by a quantum instrument acting on the measured degrees of freedom and an effectively irreversible amplification chain (cf.\ decoherence-based accounts, e.g.,~\cite{Schlosshauer2007}). The conceptual novelty of this paper is therefore not a new measurement theory, but a reframing of Bell-nonclassicality and device-independent certification in the language of time-stamping and synchronization.}

\revG{Relatedly, the term ``clock'' is used here in an \emph{event-based} operational sense: a clock is any reproducible physical procedure that outputs a time-ordered record of ticks. The entangled clock is thus closer in spirit to a stochastic ``tick generator'' (a point process with correlations) than to a self-contained harmonic oscillator. In particular, no claim is made that it competes with state-of-the-art atomic clocks in precision or stability; rather, it provides an informational synchronization primitive and (in the multi-setting regime) a device-independent privacy and unpredictability certificate for the tick record.}

\revG{For readers of \emph{Entropy}, it is useful to note the connection to Madjid and Myers' paper ``Clocks without `Time' in Entangled-State Experiments''~\cite{MadjidMyers2020}. That work emphasizes how clocks are actually used in entanglement experiments via time tags, record exchange, and post-processing, and it discusses synchronization concepts (including relativity-facing issues) without requiring a global time parameter. The present paper is complementary: it focuses on a specific operational ``tick'' convention and studies the resulting coincidence-tick rate as an informational quantity; moreover, it connects the multi-setting regime explicitly to Bell [e.g., Clauser-Horne-Shimony-Holt (CHSH)] certification and proposes ``Certified Private Time'' as a device-independent statement about the unpredictability/privacy of the time-stamp record.}

In this paper, we unify these perspectives to propose the concept of an \emph{Entangled Clock}. We define the ``tick'' of a clock operationally as the registration of a measurement outcome. \rev{Crucially, we separate (i) the \emph{local} tick streams, whose rates are fixed by the unbiased marginals, from (ii) a \revA{\emph{relational}} (coincidence-tick) stream, obtained only after Alice and Bob compare records and retain synchronized events in a designated coincidence channel.} We show that the synchronization rate of two such clocks is not merely a function of relativistic geometry, but of the information-theoretic constraints on the system. By comparing the quantum prediction against Peres' classical linear model, we identify a regime where the quantum clock \rev{exhibits a higher \emph{identified coincidence-tick rate} than this natural classical benchmark} and, when correlations at several angles are taken into account, \rev{is incompatible with any single local-realistic model reproducing all contexts simultaneously (according to Bell-type tests).}

\section{The Entangled Clock Protocol}

We construct a distributed time standard using a sequence of bipartite quantum systems. Let a source emit pairs of spin-1/2 particles (or polarization-entangled photons) in the singlet Bell state $\ket{\psi^-}$:
\begin{equation}
\ket{\psi^-} = \frac{1}{\sqrt{2}} \left( \ket{+}_A \ket{-}_B - \ket{-}_A \ket{+}_B \right),
\end{equation}
where $A$ and $B$ denote two spatially separated observers (Alice and Bob). Each observer possesses a measurement apparatus oriented along directions defined by unit vectors $\vec{a}$ and $\vec{b}$, respectively.

\subsection{Operational Time and Zeilinger's Principle}

We define operational time $T$ on each side not as a continuous background parameter, but as the accumulation of discrete events. Specifically, whenever a detector records a specific outcome (say, ``up'' or $+1$), the local clock increments by one unit:
\begin{equation}
T_i \to T_i + 1 \quad \text{if } \sigma_i = +1, \quad i \in \{A, B\}.
\end{equation}
If the outcome is ``down'' ($-1$), the clock implies a ``pause'' or a non-tick for that interval.

\rev{This choice (counting only $+1$ as a local tick) is a convention adopted for definiteness and algebraic simplicity. Other conventions are possible---for instance, counting both outcomes as ticks would make each local clock tick on every trial and would remove any ``speed'' language at the single-stream level. The nonclassical content we discuss resides in \emph{joint} (coincidence) structure, not in an absolute increase of local event rates.}

According to Zeilinger's Foundational Principle~\cite{zeil-99}, an elementary system represents the truth value of a single proposition (1 bit). For a composite system of two spin-1/2 particles, the total information capacity is 2 bits. In the singlet state $\ket{\psi^-}$, these 2 bits are entirely exhausted by the joint propositions (that is, ``spins are opposite along $z$'' and ``spins are opposite along $x$''). Consequently, the system contains \emph{zero} information about individual properties along any specific direction $\vec{n}$.
\begin{equation}
P(+)_{\vec{n}} = P(-)_{\vec{n}} = \frac{1}{2}.
\end{equation}
Thus, the \emph{local} proper time measured by Alice or Bob flows stochastically. Over a large ensemble of $N$ pairs, the local clock will record approximately $N/2$ ticks. \rev{In particular, the \emph{local} tick rate is fixed at $1/2$ per trial for both the quantum singlet and for unbiased classical models; the comparison in this paper concerns \emph{coincidence ticks} (below), not local rates.} This randomness is not merely epistemic; it is ontologically necessary because the individual particle carries no answer to the question ``are you up or down?'' until the measurement is performed. \revA{It is precisely this absence of individual definiteness that makes the clock's temporal content irreducibly \emph{relational}: meaningful time structure exists only in the joint record.}

\subsection{\rev{Coincidence Ticks and Record Exchange (Protocol Step)}}
\label{sec:coincidence-ticks}
\rev{The ``entangled clock'' figure of merit studied here is a \emph{jointly defined} tick stream. Operationally, Alice and Bob each produce a local record
\[
\mathcal{L}_A=\{(t_k^A, x_k^A, a_k)\}_{k=1}^N,\qquad \mathcal{L}_B=\{(t_k^B, x_k^B, b_k)\}_{k=1}^N,
\]
where $x_k^{A,B}\in\{+1,-1\}$ are outcomes, $a_k,b_k$ are settings, and $t_k^{A,B}$ are local time tags. They then \emph{classically exchange} (possibly after the run, offline) sufficient information to (i) match pairs within a coincidence window and (ii) identify which local $+1$ events were part of a designated coincidence channel (that is, $++$) rather than ``wrong plusses'' belonging to $+-$ or $-+$. Only after this comparison does the \emph{coincidence-tick} stream become defined. This is not ``nonlocal'' in the superluminal sense; it is a standard classical postprocessing step required to operationalize coincidences.}

\revA{In the conceptual language introduced above, the coincidence-tick stream is the operational realization of the \emph{relational time} encoded in the entangled state: it is the temporal structure that emerges only when the two individual records are brought together, mirroring the fact that the singlet's information content is entirely relational.}

\subsection{\revG{Measurement operations, destructiveness, and ``clockwork'' repeatability}}

\revG{Each trial of the protocol may be viewed as the following operational pipeline:}
\begin{enumerate}
\item \revG{A source emits one entangled pair (ideally heralded, or time-tagged at the source).}
\item \revG{Alice and Bob choose settings $a_k,b_k$ (possibly at random).}
\item \revG{Each side performs a two-outcome measurement and records a classical triple $(t_k^i,x_k^i,s_k^i)$ consisting of a local time tag $t_k^i$, an outcome $x_k^i\in\{\pm1\}$, and the setting label $s_k^i\in\{a,a'\}$ or $\{b,b'\}$.}
\item \revG{After the run, they exchange sufficient classical information to identify coincidences and, if desired, to compute Bell parameters such as CHSH.}
\end{enumerate}

\revG{In photonic implementations with avalanche photodiodes, a ``click'' is indeed a destructive detection event for the photon. This does not prevent repeated ticking, because the entangled clock is not a single closed two-particle system reused indefinitely; rather, it is a \emph{repeatable preparation--measurement procedure} fed by a stream of newly generated pairs (e.g., Spontaneous Parametric Down-Conversion). In this sense it resembles other event-based clocks such as Geiger-counter clocks or pulsed/Poissonian tick generators: the repeatability resides in the ability to reproduce the preparation and measurement across many trials, not in storing energy in a single oscillator.}

\revG{One may also envision nondestructive or quantum-nondemolition variants in matter-based platforms (e.g., repeated spin measurements or cavity-QED readout), but such refinements are not required for the core conceptual point pursued here: the operational time record is the classical list of registered events, and the nonclassical feature is contained in the multi-setting correlation structure of those events.}

\subsection{Synchronized (Coincidence) Time Rate}

While local time is random, the relative time evolution between Alice and Bob is structured by entanglement. We define the \emph{synchronized coincidence-tick rate} $R(\theta)$ as the probability that both clocks tick simultaneously in a given run \rev{in the designated coincidence channel}:
\begin{equation}
R(\theta) = P_{++}(\vec{a}, \vec{b}),
\end{equation}
where $\theta$ is the angle between the measurement vectors $\vec{a}$ and $\vec{b}$.

For the singlet state, quantum mechanics predicts the joint expectation value $E_{QM}(\theta) = \braket{\sigma_{\vec{a}} \otimes \sigma_{\vec{b}}} = -\vec{a}\cdot\vec{b} = -\cos\theta$. The probability of a coincident tick ($+1, +1$) is derived as follows:
\begin{align}
R_{QM}(\theta) &= \frac{1}{4} \left( 1 + \braket{\sigma_A} + \braket{\sigma_B} + \braket{\sigma_A \sigma_B} \right) \nonumber \\
&= \frac{1}{4} \left( 1 + 0 + 0 - \cos\theta \right) \nonumber \\
&= \frac{1}{2}\sin^2\frac{\theta}{2}.
\label{eq:qm_rate}
\end{align}

\textit{Implementation Note:} While we describe the protocol using spin-1/2 particles, an equivalent implementation uses polarization-entangled photon pairs. In that case, $\ket{+}/\ket{-}$ correspond to horizontal/vertical polarizations, and the detectors are linear polarizers. A ``tick'' corresponds to a photon passing through the polarizer and triggering an avalanche photodiode. \rev{Coincidence ticks are identified by standard time-tag matching and outcome comparison.}

\section{Classical Simulation: Peres' Bomb}

To determine if the behavior of the entangled clock is unique, we must compare it to a classical standard. A robust classical analog for the singlet state was provided by Peres~\cite{peres222} and is often referred to as the ``bomb fragment'' model.

\rev{We stress at the outset: Peres' model is \emph{not} claimed to be the unique classical comparator. It is used here as a natural, isotropic local-hidden-variable (LHV) construction with unbiased marginals that obeys standard Boole--Bell constraints. For any \emph{single} fixed angle $\theta^{*}$, one may engineer contextual classical mechanisms that reproduce $P_{++}(\theta^{*})$ (cf.\ Sec.~\ref{sec:certify-quantumness}). The discriminating content arises when one demands a \emph{single} LHV model reproducing \emph{all} settings simultaneously, i.e., when one moves from a one-angle fit to the multi-setting correlation polytope.}

Consider a bomb initially at rest (total angular momentum $\vec{J}=0$) that explodes into two fragments with opposite angular momenta $\vec{J}_1$ and $\vec{J}_2 = -\vec{J}_1$. The direction of $\vec{J}_1$ is random and distributed uniformly over the sphere. The observers measure the sign of the projection of the angular momentum onto their axes $\vec{a}$ and $\vec{b}$:
\begin{equation}
r_\alpha = \sign(\vec{a} \cdot \vec{J}_1), \quad r_\beta = \sign(\vec{b} \cdot \vec{J}_2).
\end{equation}
Peres derives the correlation function for this macroscopic local hidden variable (LHV) model geometrically. The correlation depends on the overlap of the hemispheres defined by the vectors $\vec{a}$ and $\vec{b}$ on the unit sphere of possible angular momenta. The classical expectation value is strictly linear in $\theta$:
\begin{equation}
E_{cl}(\theta) = -1 + \frac{2\theta}{\pi} \quad \text{for } \theta \in [0, \pi].
\label{eq:peres_corr}
\end{equation}
The corresponding classical synchronized tick rate is:
\begin{equation}
R_{cl}(\theta) = \frac{1}{4}(1 + E_{cl}) = \frac{1}{4}\left(1 - 1 + \frac{2\theta}{\pi}\right) = \frac{\theta}{2\pi}.
\label{eq:cl_rate}
\end{equation}

\section{Comparison: When Quantum Coincidence Ticks Exceed the Classical Benchmark}
\label{sec:comparison}

\rev{In what follows, any ``faster ticking'' language refers \emph{only} to the \emph{identified coincidence-tick rate} $R(\theta)=P_{++}(\vec a,\vec b)$ under the convention that a tick corresponds to the $+1$ outcome and that the coincidence tick is the $++$ channel after record exchange. The local tick rates at Alice and Bob remain fixed at $1/2$ (unbiased marginals) for both the quantum singlet and for unbiased classical models. No claim is made here about improved metrological precision, stability, or Allan deviation of a physical oscillator.}

We now compare the two synchronization rates. We can distinguish three ``cardinal'' regimes where the physics appears identical, and one ``anomalous'' regime where the quantum clock diverges.

\subsection{The Cardinal Regimes}
\begin{enumerate}
    \item \textbf{Perfect Anti-Synchrony ($\theta=0$):} Detectors are aligned ($\vec{a} = \vec{b}$).
    \[
    R_{QM}(0) = 0, \quad R_{cl}(0) = 0.
    \]
    Because $\ket{\psi^-}$ is perfectly anti-correlated (singlet), if Alice's clock ticks, Bob's does not. The clocks are perfectly out of phase.

    \item \textbf{Non-Correlation ($\theta=\pi/2$):} Detectors are orthogonal ($\vec{a} \perp \vec{b}$).
    \[
    R_{QM}(\pi/2) = 1/4, \quad R_{cl}(\pi/2) = 1/4.
    \]
    Here, $E_{QM} = 0$ and $E_{cl} = 0$. The clocks are statistically independent.

    \item \textbf{Perfect Synchrony ($\theta=\pi$):} Detectors are anti-aligned ($\vec{a} = -\vec{b}$).
    \[
    R_{QM}(\pi) = 1/2, \quad R_{cl}(\pi) = 1/2.
    \]
    Since $\ket{\psi^-}$ implies opposite spins, measuring opposite directions yields identical results ($++$ or $--$). The clocks tick together $50\%$ of the time.
\end{enumerate}
In these regimes, the informational constraint of the quantum bit yields the same statistics as the classical angular momentum conservation.

\subsection{The Anomalous Regime}

The divergence between the single-unit universe governed by quantum information and one governed by classical statistical mechanics appears at intermediate angles. We define the \emph{synchronization excess} $\Delta(\theta)$ as:
\begin{equation}
\Delta(\theta) \equiv R_{QM}(\theta) - R_{cl}(\theta) = \frac{1}{2}\sin^2\frac{\theta}{2} - \frac{\theta}{2\pi}.
\end{equation}
To find the maximum deviation, we find the extrema by setting $d\Delta/d\theta = 0$:
\begin{equation}
\frac{1}{4}\sin\theta - \frac{1}{2\pi} = 0 \implies \sin\theta = \frac{2}{\pi}.
\end{equation}
This yields two solutions in $[0, \pi]$:
\begin{align}
\theta_1 &= \arcsin(2/\pi) \approx 0.69~\text{rad}~(39.5^\circ), \quad (\Delta < 0), \\
\theta_2 &= \pi - \theta_1 \approx 2.45~\text{rad}~(140.5^\circ), \quad (\Delta > 0).
\end{align}

At $\theta_1 \approx 39.5^\circ$, the classical correlation is stronger (more negative), so the quantum clock ticks \emph{less} frequently than the classical prediction. This is the regime where the classical linear correlation drops faster than the quantum cosine.

However, at $\theta_2 \approx 140.5^\circ$, the quantum clock exhibits a significant excess of coincidence ticks (see Fig.~\ref{fig:rates}).
\begin{itemize}
    \item Classical Rate: $R_{cl}(140.5^\circ) \approx 0.390$.
    \item Quantum Rate: $R_{QM}(140.5^\circ) \approx 0.443$.
\end{itemize}
\rev{The relative excess at this angle is}
\[
\rev{\frac{R_{QM}-R_{cl}}{R_{cl}}\bigg|_{\theta\approx 140.5^\circ}\approx \frac{0.443-0.390}{0.390}\approx 0.136.}
\]

\begin{figure}[t]
\centering
\begin{tikzpicture}
\begin{axis}[
    width=\columnwidth,
    height=0.7\columnwidth,
    xlabel={$\theta$ (radians)},
    ylabel={Synchronization Rate $R(\theta)$},
    xmin=0, xmax=3.14159,
    ymin=0, ymax=0.55,
    xtick={0, 0.785, 1.571, 2.356, 3.14159},
    xticklabels={$0$, $\frac{\pi}{4}$, $\frac{\pi}{2}$, $\frac{3\pi}{4}$, $\pi$},
    legend pos=north west,
    grid=major,
    grid style={dashed, gray!30},
]
% Quantum rate
\addplot[blue, thick, domain=0:pi, samples=100]
    {0.5*sin(deg(x/2))^2};
\addlegendentry{$R_{QM}(\theta)$}

% Classical rate
\addplot[red, thick, dashed, domain=0:pi]
    {x/(2*pi)};
\addlegendentry{$R_{cl}(\theta)$}

% Mark the critical points
\addplot[only marks, mark=*, mark size=2.5pt, black]
    coordinates {(2.4515, 0.443) (2.4515, 0.390)};

% Labels for markers
\node[anchor=west, font=\scriptsize] at (axis cs:2.5, 0.443) {QM};
\node[anchor=west, font=\scriptsize] at (axis cs:2.5, 0.390) {cl};

% Vertical line at theta_2
\draw[dotted, thick] (axis cs:2.4515,0) -- (axis cs:2.4515,0.5);
\node at (axis cs:2.4515,0.52) [anchor=south] {$\theta_2$};

\end{axis}
\end{tikzpicture}
\caption{\rev{Comparison of the quantum ($R_{QM}$, solid blue) and Peres-benchmark classical ($R_{cl}$, dashed red) $++$ coincidence-tick rates as functions of the relative detector angle $\theta$. The curves coincide at $\theta = 0, \pi/2, \pi$ but diverge maximally at $\theta_2 \approx 140.5^\circ$, where the \emph{identified coincidence-tick rate} predicted by quantum theory exceeds this natural isotropic LHV benchmark.}}
\label{fig:rates}
\end{figure}

The cosine modulation of the quantum correlation function, which is consistent with Zeilinger's information-invariance principle, permits a higher density of coincident events than the linear constraint imposed by the isotropic bomb-fragment model. In particular, at $\theta_2$ the quantum prediction exceeds the linear classical benchmark by about $13.6\%$.
Figure~\ref{fig:delta} explicitly shows this synchronization excess $\Delta(\theta)$. \rev{We emphasize again: this is a statement about the $++$ coincidence channel under an agreed operational convention, not about an absolute increase of local ticking.} The positive lobe represents a regime where, if one demands a single local realistic model that applies to \emph{all} angles simultaneously, the quantum predictions force the clocks to agree more often than such a model can allow.

\begin{figure}[t]
\centering
\begin{tikzpicture}
\begin{axis}[
    width=\columnwidth,
    height=0.5\columnwidth,
    xlabel={$\theta$ (radians)},
    ylabel={$\Delta(\theta)$},
    xmin=0, xmax=3.14159,
    ymin=-0.06, ymax=0.06,
    xtick={0, 0.785, 1.571, 2.356, 3.14159},
    xticklabels={$0$, $\frac{\pi}{4}$, $\frac{\pi}{2}$, $\frac{3\pi}{4}$, $\pi$},
    grid=major,
    grid style={dashed, gray!30},
]
\addplot[blue, thick, domain=0:pi, samples=100]
    {0.5*sin(deg(x/2))^2 - x/(2*pi)};
\draw[dashed, black] (axis cs:0,0) -- (axis cs:pi,0);
\addplot[only marks, mark=*, mark size=2pt, red]
    coordinates {(0.69, -0.053) (2.4515, 0.053)};
\node at (axis cs:0.8, -0.04) [anchor=west, font=\scriptsize] {$\theta_1$};
\node at (axis cs:2.35, 0.04) [anchor=east, font=\scriptsize] {$\theta_2$};
\end{axis}
\end{tikzpicture}
\caption{The synchronization excess $\Delta(\theta) = R_{QM} - R_{cl}$. The quantum clock lags at acute angles ($\theta_1 \approx 39.5^\circ$) but leads at obtuse angles ($\theta_2 \approx 140.5^\circ$).}
\label{fig:delta}
\end{figure}

\section{Contextuality, Polytopes, and Unperformed Experiments}

Why does the quantum clock \rev{exceed the Peres benchmark in the $++$ coincidence channel for obtuse angles}? The answer lies in the nature of the information carried by the system and the role of contextuality.

In Peres' model~\cite{peres222}, the angular momentum vector $\vec{J}$ exists independently of the measurement. Peres argues that Bell's inequality can be derived by constructing a table of outcomes for experiments actually performed (settings $\vec{a}, \vec{b}$) and hypothetical experiments \emph{not} performed (settings $\vec{a}', \vec{b}'$). For a classical system, this table must be internally consistent because the value $\vec{J}$ pre-exists and determines the outcome for any possible question. In the specific bomb-fragment model, the overlap of the corresponding hemispheres on the unit sphere yields a strictly linear correlation function, Eq.~\eqref{eq:peres_corr}, which in turn strictly limits the synchronized tick rate.

In the quantum case, following Zeilinger~\cite{zeil-99}, the bipartite system carries only 2 bits of information, which are already exhausted by the specific joint state. There is no ``spare'' information to encode definite answers to all possible, mutually incompatible measurement questions along directions $\vec{a}',\vec{b}'$. From this information-theoretic perspective, the system is not constrained to maintain consistency across a full table of actual and counterfactual outcomes, and thus is not subject to the Boole--Bell-type constraints~\cite{froissart-81,pitowsky-86} that enforce the existence of a single joint probability distribution for all settings and, in particular, lead to linear behavior in models like Peres' bomb. This ``freedom from counterfactual definiteness'' is one way to understand why the quantum correlation $E_{QM}(\theta)$ follows the cosine law and exceeds the linear bomb-model bound at $\theta \approx 140^\circ$.

\rev{Equivalently (and more geometrically), the set of correlation families compatible with a single LHV model forms a convex polytope in the space of multi-setting correlations~\cite{froissart-81,pitowsky-86}. A one-angle agreement is easy to engineer; what is constrained is the \emph{simultaneous} reproduction of several angles by one underlying joint distribution. The ``cosine-vs-linear'' mismatch visible already at the level of Fig.~\ref{fig:rates} is one slice of this broader polytope structure, which is operationalized by Bell-type inequalities.}

\subsection{Certifying Quantumness}
\label{sec:certify-quantumness}

While the rate $R_{QM} \approx 0.44$ is notably higher than the standard classical rate $R_{cl} \approx 0.39$, one must be cautious. A single measurement setting cannot distinguish quantum from classical clocks. As noted in recent work on contextuality~\cite{svozil-2022-epr}, for any \emph{single} fixed angle $\theta^*$, one can construct a local classical model (e.g., Aerts' ``broken elastic band'' model~\cite{aerts-69}) that reproduces the quantum probability $R_{QM}(\theta^*)$ exactly. In such models, the hidden variable space is deformed or the probability density is non-uniform to mimic the cosine statistics locally.

The signature of quantum synchronization is \emph{contextual}: the same underlying state produces the \emph{entire} cosine curve across all angles simultaneously. To certify quantumness, Alice and Bob must therefore vary their angles and compare several correlations at once. Any local realistic model must admit a single joint probability distribution that reproduces all these correlations and thus obeys Bell-type inequalities, whereas the quantum clock yields a family of cosine values that violates these bounds. The enhanced synchronized tick rate in the obtuse regime is then seen not as an isolated anomaly, but as one facet of a globally nonclassical correlation structure.

The excess synchronization rate $\Delta(\theta) > 0$ in the obtuse regime is directly related to violations of Bell-type inequalities such as the Clauser--Horne--Shimony--Holt (CHSH) inequality~\cite{chsh}.
The CHSH parameter $S$ is composed of four correlation terms:
\begin{equation}
S = |E(\vec{a},\vec{b}) - E(\vec{a},\vec{b}') + E(\vec{a}',\vec{b}) + E(\vec{a}',\vec{b}')|.
\end{equation}
The classical bound $S \le 2$, obtained from the convex hull computation of the corresponding correlation polytope~\cite{froissart-81,pitowsky-86}, arises from the requirement that all four correlations be marginals of a single joint probability distribution, that is, from the assumption that the table of actual and counterfactual outcomes is internally consistent.
The linear correlation function~\eqref{eq:peres_corr} is one explicit realization of such a local realistic model.
The quantum violation (approaching the Tsirelson bound $2\sqrt{2}$) and the enhanced synchronization rate in the obtuse regime are two manifestations of the same underlying phenomenon.

\rev{\subsection{Explicit CHSH violation from the obtuse-angle excess}}
\rev{To make the connection concrete, choose coplanar analyzer directions
specified by angles (in radians the violation is maximized as long as the following conditions are met: (i)  Alice's two settings are orthogonal ($\pi/2$ apart); (ii)  Bob's two settings are orthogonal ($\pi/2$ apart); and (iii) Bob's basis is rotated by $\pi/4$ relative to Alice's basis)
\[
a=0,\qquad a'=\frac{\pi}{2},\qquad b=\frac{\pi}{4},\qquad b'=\frac{3\pi}{4}.
\]
For the singlet, $E_{QM}(x,y)=-\cos(x-y)$, hence
\begin{align}
E(a,b)&=-\cos\frac{\pi}{4}=-\frac{\sqrt2}{2},
\nonumber \\
E(a,b')&=-\cos\!\Big(\!-\frac{3\pi}{4}\Big)=+\frac{\sqrt2}{2},
\nonumber \\
E(a',b)&=-\cos\frac{\pi}{4}=-\frac{\sqrt2}{2},
\nonumber \\
E(a',b')&=-\cos\!\Big(\!-\frac{\pi}{4}\Big)=-\frac{\sqrt2}{2}. \nonumber
\end{align}
Therefore
\[
S_{QM}=\left|-\frac{\sqrt2}{2}-\frac{\sqrt2}{2}-\frac{\sqrt2}{2}-\frac{\sqrt2}{2}\right|=2\sqrt2>2.
\]
This standard choice includes an obtuse relative angle of $|a-b'|=3\pi/4=135^\circ$, which lies in the positive lobe of Fig.~\ref{fig:delta} (indeed $\Delta(3\pi/4)>0$). Thus, the same cosine-vs-linear mismatch that yields an excess $++$ coincidence-tick rate at obtuse angles also underlies the familiar CHSH violation.}

\section{Experimental Considerations}

Implementing this protocol requires addressing practical constraints. In an ideal scenario, every particle pair results in a potential tick. In reality, photon detectors have finite efficiency $\eta < 1$.

If $\eta < 1$, the clocks will frequently ``miss'' ticks even when the quantum outcome would have been $+1$. However, the synchronization rate $R(\theta)$ is typically measured as a coincidence rate normalized by the total number of emitted pairs (or inferred from singles counts). If we define the measured rate $R_{exp}(\theta)$, we have:
\begin{equation}
R_{exp}(\theta) \approx \eta_A \eta_B R_{QM}(\theta).
\end{equation}
This reduced efficiency lowers the absolute rate of ticking for both classical and quantum models. The crucial comparison is the \emph{relative} rate or the visibility of the interference fringe. A high-visibility experiment (using, e.g., spontaneous parametric down-conversion and superconducting nanowire single-photon detectors with $\eta > 0.9$) would clearly resolve the difference between the linear classical prediction and the sinusoidal quantum prediction.

Furthermore, the definition of ``simultaneity'' requires a coincidence window $\tau_c$. If the path lengths to Alice and Bob differ, or if there is electronic jitter, ticks might not register as simultaneous. Standard quantum optical techniques allow $\tau_c$ to be on the order of nanoseconds. \rev{As emphasized in Sec.~\ref{sec:coincidence-ticks}, the coincidence-tick stream is obtained only after classical record exchange (time tags, outcomes, and in the multi-setting case also the settings) and coincidence matching.} The ``Entangled Clock'' is thus a statistical construct: it does not provide a continuous readout for a single run but establishes a time scale over an ensemble of events, where the density of synchronized events serves as the metrological standard.

\rev{We note again that this notion of ``metrological standard'' is meant here in the operational and informational sense of a reproducible \emph{rate of identified coincidence events}, not as a claim of superior oscillator performance.}

\revG{From the perspective of physical implementation, the relevant ``system'' is therefore intrinsically open: it includes the source (pair generation), transmission channels, detectors, and the classical time-tagging electronics. The protocol does not assume coherent unitary evolution of a single isolated clockwork between ticks; instead, it assumes repeatable preparation and measurement producing a stable classical event record. This places the construction closer to the operational practice of time-tagged entanglement experiments (where post-processing defines coincidence structure) than to autonomous oscillator clocks.}

\section{Certified Private Time}
\label{sec:certified-private-time}

The discussion above highlights a crucial distinction: while a classical model can be engineered to reproduce quantum statistics for a single context, it fails to reproduce the full correlation structure across multiple contexts. This brings us to a stronger application of the entangled clock: the certification of ``private time.''

As reviewed by Ac\'in and Masanes~\cite{Acin2016CertifiedRandomness}, the violation of Bell inequalities serves as a device-independent certificate for randomness. In the context of random number generation (QRNG), this certification ensures that the output bits were not pre-determined by any hidden variable $\lambda$ available to an adversary (Eve). \rev{Here we adapt the same logic to a \emph{time-stamp record}: what is certified is the unpredictability (relative to side information) of the ordered sequence of identified coincidence ticks, under the standard device-independent assumptions (in particular, no adversarial foreknowledge of future settings).}

\revG{In particular, ``certification'' here has the same logical status as in device-independent randomness: it is a statement about the incompatibility of the observed input--output statistics with pre-programmed local-hidden-variable strategies, given standard assumptions (that is, measurement independence and no information leakage of future settings). It is not a claim to have derived a fundamental solution of the measurement problem, nor a claim that the devices implement a closed dynamical time observable.}

\subsection{The Memory-Stick Clock}
Consider the temporal equivalent of the ``memory-stick attack'' described in Ref.~\cite{Acin2016CertifiedRandomness}. An adversary, Eve, manufactures two clocks and gives them to Alice and Bob. These clocks are classical devices containing a pre-recorded sequence of ticks derived from a hidden variable $\lambda$. To Alice and Bob, the ticks appear stochastic (e.g., mimicking radioactive decay), and they may even exhibit correlations. However, because the sequence is pre-determined, Eve possesses a ``lookup table'' for time. She knows exactly when Alice's clock will tick before it happens. In such a scenario, Alice's time is not her own; it is fully correlated with the adversary's timeline.

In a classical universe (or one governed by local hidden variables), one cannot prove that a clock is not a playback device. The flow of time could be merely the reading of a script written at the Big Bang (super-determinism aside).

\subsection{Protocol for Time Certification}
The entangled clock allows Alice and Bob to rule out this ``playback'' scenario. By switching between measurement settings (angles $\vec{a}, \vec{a}'$ for Alice and $\vec{b}, \vec{b}'$ for Bob) and accumulating the statistics of their coincident ticks, they can estimate the CHSH parameter $S$.

If they observe $S > 2$, they certify that the joint state does not admit a local realistic description. Physically, this implies that the specific outcomes (the ticks) did not exist prior to the measurement events.
Consequently, the time signal generated by the clock is:
\begin{enumerate}
    \item \textbf{Intrinsic:} The ticks are generated \emph{in situ} by the act of measurement, not read from a memory.
    \item \textbf{Private:} \rev{the coincidence-tick record is unpredictable to any adversary whose side information is not correlated with the (future) setting choices, in the same device-independent sense as certified randomness. In particular, the certification concerns the \emph{ordering/time-stamping record} of identified coincidence ticks, not the performance of a physical oscillator.}
\end{enumerate}

\rev{\subsection{Cryptographic vs.\ metrological meaning of ``private time''}}
\rev{The intended meaning of ``private time'' is cryptographic and informational: Alice and Bob can maintain a time-stamped record (a sequence of coincidence ticks with associated time tags) whose fine-grained content cannot be reconstructed by an external observer without access to the settings and without violating the device-independent constraints implied by $S>2$. This should not be confused with a metrological claim such as improved phase resolution, reduced timing jitter, or improved Allan deviation.}

\subsection{Escaping the Block Universe}
This leads to a radical reinterpretation of the ``flow'' of time. In a standard operational view, time is a parameter inferred from correlations between a clock system and the rest of the universe. However, if the clock's ticks are certified random via Bell violation, the clock is dynamically decoupled from the rest of the universe until the moment of the tick.

A ``Certified Private Clock'' thus ticks at a rate that is not only statistically distinct \rev{at the level of the $++$ coincidence-tick benchmark comparison} (the ``faster'' rate at obtuse angles discussed in Sec.~\ref{sec:comparison}) but ontologically distinct. It generates a local history that no external observer could have predicted. Just as device-independent protocols use Bell violation to expand a short private random seed into a distinct, private key, the entangled clock expands a short interval of private operational agency (the choice of settings) into a sustained, private flow of time.

This certification is robust against the ``middleman attack'' often cited in foundational debates. Even if Eve builds the source and controls the channels, she cannot force the devices to violate the CHSH inequality with pre-programmed ticks unless she knows the measurement settings (the context) in advance. To prevent this, Alice and Bob must ensure ``measurement (parameter)independence''~\cite{shimony_1993}. \rev{This is the same assumption used in standard device-independent QRNG and QKD; no stronger assumption is invoked here.} This can be achieved either through genuine free choice or, alternatively, by consuming a short, predefined private random key to govern the settings. In the latter case, the system functions as a temporal implementation of Ac\'in's expansion protocol~\cite{Acin2016CertifiedRandomness}, growing the initial finite seed into a sustained, sovereign timeline that is mathematically guaranteed to be uncorrelated with any master clock Eve might hold.

\section{Conclusion}

We have proposed an operational protocol for \emph{Certified Private Time}, in which spatially separated quantum clocks are synchronized via singlet-state correlations. By adapting device-independent certification methods~\cite{Acin2016CertifiedRandomness}, we showed that if observers consume a private seed to ensure measurement independence, the resulting timeline is certified to be ``sovereign''---generated \emph{in situ} and mathematically uncorrelated with any pre-existing external script.

Crucially, \rev{the nontrivial ``rate'' comparison in this work concerns a \emph{jointly defined coincidence-tick stream} (that is, $++$) obtained after classical record exchange and coincidence identification. Locally, the marginal tick rates remain fixed by unbiased statistics.} For instance, at obtuse relative angles ($\theta \approx 140^\circ$), the synchronized $++$ coincidence-tick rate exhibits a ``temporal excess'' of ${\sim}13.6\%$ relative to Peres' isotropic bomb-fragment model. \rev{This does not constitute an absolute claim that quantum clocks ``run faster'' in the usual metrological sense; rather, it is an operational statement about the density of identified synchronized events under a specified convention.}

This deviation is a direct signature of contextuality. Unlike the classical model, the quantum system is not constrained to maintain consistency with unperformed measurements. The violation of Bell-type inequalities thus serves a dual purpose: it certifies the privacy of the time stream against a middleman adversary, and it confirms that the \rev{multi-setting} correlation structure cannot be embedded into any single local-realistic model. \rev{In this sense, the obtuse-angle ``excess'' lobe and the CHSH violation are two views of the same nonclassical resource.}

\revA{We close by returning to the conceptual core. The time defined by the entangled clock is \emph{relational} in a precise sense: because the singlet state encodes no definite properties of either subsystem individually, the only meaningful temporal structure is contained in the joint record---the coincidence-tick stream that emerges when Alice's and Bob's local records are brought together. This relational character has no single-party analogue and no classical counterpart that could simultaneously reproduce the full angular dependence of the coincidence rate. In the multi-setting (Bell-violating) regime, this relational time furthermore acquires a device-independent privacy guarantee: the joint tick record is certified to be unpredictable to any adversary who lacks foreknowledge of the measurement settings. The certified private time is thus irreducibly relational---it exists only in the ``between,'' not in either party alone.}

\revG{Finally, we contrast this construction with practical atomic clocks. Atomic clocks are engineered oscillators stabilized by coherent interrogation and feedback, and their performance is judged by frequency stability/accuracy metrics. The present entangled-clock protocol is different in kind: it produces an event-based time-stamp record whose \emph{joint} coincidence structure can exceed a natural isotropic LHV benchmark at certain angles, and whose \emph{privacy and unpredictability} can be device-independently certified in the multi-setting (Bell-violating) regime.}

\begin{acknowledgments}
This text was partially created and revised with assistance from large language models. All content, ideas, and prompts were provided by the author.

The entangled clock metaphor was suggested by Adrienne Toman during a hike to the Michelberg.

This research was funded in whole or in part by the Austrian Science Fund (FWF), grant number DOI: 10.55776/PIN5424624. The author acknowledges TU Wien Bibliothek for financial support through its Open Access Funding Programme.

\textbf{Data Availability:} No new data were created or analyzed in this study. Data sharing is not applicable to this article.

\textbf{Conflicts of Interest:} The author declares no conflict of interest.

\end{acknowledgments}

\bibliography{svozil}

\end{document}